
\overfullrule 0pt
\input harvmac

\def \s{\sigma}
\def \b{\beta}
\def \a{\alpha}
\def \g{\gamma}
\def \d{\delta}
\def \e{\epsilon}
\def \l{\lambda}

\def \ph{\phi}

\def \Ph{\Phi}

\def \z {\zeta}
\def \m{\mu}

\def\npb{{Nucl.\ Phys.\ }{\bf B}}
\def\physrep{Phys.\ Reports\ }
\def\plb{{Phys.\ Lett.\ }{\bf B}}

\def\prl{Phys.\ Rev.\ Lett.\ }
\def\ptp{Prog.\ Th.\ Phys.\ }

\def\zpc{Z.\ Phys.\ {\bf C}}

\def\dM{M^*}
\def\tm{\tilde m}
\def\q{\quad}
\def\psib{\overline{\psi}}
\def\lambdab{\overline{\lambda}}
\def\sy{supersymmetry}
\def\sic{supersymmetric}

\def\ssm{supersymmetric standard model}
\def\sm{standard model}

\def\lf{16\pi^2}
\def\llf{(16\pi^2)^2}

\Title{LTH 334}{Soft supersymmetry breaking and finiteness}
\centerline{I. Jack and D. R. T. Jones}
\bigskip
\centerline{\it DAMTP, University of Liverpool, Liverpool L69 3BX, U.K.}
\vskip .3in
We calculate the two-loop $\beta$-functions for soft supersymmetry-breaking
interactions in a general \sic\ gauge theory, emphasising the role of
the evanescent couplings and masses that are
generic to a non-supersymmetric theory. We also
show that a simple set of conditions sufficient for one-loop finiteness
renders a theory two-loop finite, and we speculate on the possible
significance of finite softly broken theories.

\Date{May 1994}

Recently we addressed the issue of the applicability of regularisation
by dimensional reduction (DRED)
\ref\dred{W. Siegel, \plb84 (1979) 193.}
\ref\cjn{D. M. Capper, D. R. T. Jones and P. van
Nieuwenhuizen, \npb167 (1980) 479.}
to non-\sic\ theories\ref\jjra
{I.~Jack, D.~R.~T.~Jones and
K.~L.~Roberts, \zpc62 (1994) 161.}
\ref\jjrb{I.~Jack, D.~R.~T.~Jones and
K.~L.~Roberts,
Liverpool preprint LTH 329 (Z. Phys. C, to be published).}.

The  conclusion of this analysis was that DRED is indeed a consistent
procedure. Moreover, for calculations performed at a given choice of
renormalisation scale $\mu$, the usual procedure of ignoring the
``evanescent couplings'' (for an account of these see below or Refs.~\jjra,
\jjrb) is perfectly valid. However we indicated that care must be exercised
if the object is to relate calculations at two different values of $\mu$
 by means of appropriate $\b$-functions.

DRED has been used in non-\sic\ theories because certain calculations
are simplified if the Dirac matrix algebra can be carried out in 4
rather  than $d$ dimensions. Another class of theories where use of DRED
  rather than DREG (conventional dimensional regularisation) is  clearly
indicated is softly broken \sic\ theories.  Here DRED  is preferable to
DREG  because the dimensionless couplings in a  softly broken \sic\
theory renormalise (using minimal subtraction)  exactly as in the
corresponding \sic\ one.

The \ssm\ consists of the minimal \sic\ extension of the \sm\ plus all possible
soft-breaking terms. A popular constraint on the large number of parameters
thus invoked is the assumption of a simple form for the soft-breaking terms
at a scale   corresponding to or near the Planck mass. It is then
necessary  to evolve the parameters using their $\beta$-functions
to determine the low energy theory. The one loop $\beta$-functions
have been known for some time\ref\ikks{K. Inoue, A. Kakuto,  H. Komatsu
and S. Takeshita \ptp 68 (1982) 927; erratum {\it ibid\/} 70 (1983) 330.},
and the two-loop $\b$-functions have been calculated recently
for a general theory\ref\yama{Y. Yamada, \plb316 (1993) 109;
\prl72 (1994) 25.}\nref\mva{S. P. Martin and M. T. Vaughn,
\plb318 (1993) 331.}\nref\mv{S. P. Martin and
M. T. Vaughn, Northeastern preprint NUB-3081-93TH.}--\ref\yam{Y.
Yamada, KEK preprint 93-182.}.

The method used in Refs.~\mva, \mv\ is somewhat different to that used in Refs.
\yama, \yam.  The calculations of Refs.~\yama, \yam\ are performed using
the superfield formalism together with DRED,  the \sy\ breaking terms
being managed by the use of spurion superfields. The authors of Refs.
\mva, \mv\ work using component fields, and they start with results obtained
using dimensional  regularisation, obtaining the results corresponding
to DRED by making a coupling constant redefinition,
after which their results agreewith those of Refs.~\yama, \yam.
We have independently
calculated the same  set of $\b$-functions, obtaining a slightly
different result. As we shall  explain below, we ascribe this difference
to our treatment of the  evanescent $\e$-scalar mass.

The difference between DRED and DREG resides in the treatment of the
gauge fields as one passes from 4 to $4-\e$ dimensions in
order to regulate the theory. In DREG the number of
gauge fields is simply continued to $4-\e$, while in DRED
the theory in $4-\e$ dimensions is obtained by compactification: the number of
fields is unaltered, but $\e$ components of a vector multiplet
become scalars, called ``$\e$-scalars''\cjn. It can be shown that the
$\b$-functions obtained by DRED (when properly defined--see later)
can be obtained by redefinition of
coupling constants from the results for DREG--indeed, this is a crucial
requirement for the consistency of DRED\jjra\jjrb
\ref\hvd{R. van Damme and G. 't Hooft, \plb150 (1985) 133.}
\ref\ij{I. Jack, \plb147 (1984) 405\semi
G. Curci and G. Paffuti, \plb148 (1984) 78\semi
D. Maison, \plb150 (1985) 139.}.
In order to show this in general, it is important to
realise\jjra\jjrb\hvd\
that the terms in the Lagrangian involving $\e$-scalars will not in
general renormalise in the same way as the terms involving the
corresponding vector fields, since they are not related by $d$-dimensional
Lorentz invariance. We should introduce new couplings--called
``evanescent''--for the terms with $\e$-scalars.
However, for a supersymmetric theory the evanescent couplings are
nevertheless related by supersymmetry to the corresponding real ones,
and so in practice may be identified with them\ij. To put it another way,
if the evanescent couplings are set equal to their ``natural'' \sic\ values at
one value of the renormalisation scale $\mu$, then they remain equal at all
values of $\mu$.  For a
non-supersymmetric theory, on the other hand, it is crucial to distinguish the
evanescent couplings from the real ones.
Moreover, one must
be careful to include all the evanescent couplings which are allowed--not
only those which appear upon compactification of the original 4-dimensional
theory--since they will in general be generated in the renormalisation
process. The softly broken $N=1$
supersymmetric theory under discussion is a case in point. Here the
dimensionless evanescent couplings may still be identified with the
supersymmetric gauge couplings, since their evolution as a function of $\mu$ is
not affected by the
soft-breaking terms. Furthermore,
the compactification to $4-\e$ dimensions does not generate
any dimensionful $\e$-scalar couplings. Nevertheless, there are, for instance,
one-loop diagrams generating divergent contributions to the $\e$-scalar mass,
and hence to discuss the renormalisation of the theory using DRED,
it behoves us to
include an $\e$-scalar mass term. This is essential in order to obtain the
$\b$-functions for DRED from those for DREG by coupling constant redefinition,
as was realised by the authors of Ref.~\mv. However, at the end of their
calculations they set the $\e$-scalar mass to zero.

This poses an obvious problem, as follows. In a previous paper, we
demonstrated that it is perfectly possible to assign arbitrary values
to the evanescent couplings for calculations
performed using DRED at a given momentum
scale in a non-\sic\ theory; and so, for example, zero for the $\e$-scalar mass
is a possible and indeed the obvious choice. But it does not follow that
if one is
interested in studying the running of the couplings (in other words relating
calculations at one value at $\mu$ to those at another), one is simply
entitled to set the $\e$-scalar mass
identically zero throughout.
 Suppose  the $\e$-scalar mass was chosen  to be zero at the unification
  scale, and we want to calculate parameters at the weak scale in terms
of those  at the unification scale.  The $\e$-scalar mass $\b$-function
contains contributions  proportional to the soft scalar and gaugino
masses, and consequently the $\e$-scalar
  would develop a mass as we run down to the weak scale. Moreover,
the two-loop $\b$-function for the soft-breaking mass
$m^2$ acquires contributions from a non-zero $\e$-scalar mass within DRED.
We thus obtain at the weak scale a theory with a non-zero $\e$-scalar mass,
and a certain set of values for the physical couplings.
As we discussed in Ref.~\jjrb, this theory is
equivalent
to one with a vanishing $\e$-scalar mass, and a different set of values for the
physical couplings, at the same scale; but this is $\it not$ the same theory as
would be obtained by simply setting the $\e$-scalar mass identically zero
by fiat throughout the process of running the couplings.

We now turn to presenting the details of our calculations of the
$\b$-functions for the softly broken $N=1$ theory. These are computed
using DRED.
We should stress here
that our definition of DRED includes a particular prescription for the
construction of subtraction diagrams. In fact, two versions of
DRED have been discussed in the literature\cjn\jjra\jjrb\hvd
\ref\jo{I. Jack and H. Osborn, \npb 249 (1985) 472.},
which differ in their use of counterterms within
subtraction diagrams.
In the first version of DRED\cjn\jjra\jjrb,
counterterms are calculated by requiring that
all graphs, including those with external $\e$-scalars, are
finite--leading to evanescent couplings being assigned counterterms
different from those for physical couplings;
in the second version\hvd,
on the other hand, evanescent couplings are assigned
the same counterterms as the physical gauge coupling.
It is easy to see that
the first version corresponds to the normal process of
diagram-by-diagram subtraction, in which one subtracts diagrams in which
divergent subdiagrams are replaced by counterterm insertions with the same
pole structure. Moreover, it can be shown\jjrb\ that in general it is the
results obtained using the first version of DRED which can be obtained by the
coupling constant redefinition from those of DREG.
The results obtained using the second prescription cannot\hvd\ in general be
obtained by coupling constant redefinition from those of DREG.
In addition, it is clear, as
argued in Ref.~\hvd, that the second version of DRED will in
general lead to loss of unitarity. For a fully
supersymmetric theory the distinction between the two versions of DRED
disappears\jo;
however, in the present case of the softly broken theory
we should be more careful, and
henceforth, we shall use the term DRED to refer to the first version alone,
unless otherwise stated. It is this version of DRED which we use for our
calculations.

In computing the $\b$-functions for the softly broken $N=1$ \sic\
 theory using DRED we must
include all interactions with $\e$-scalars consistent with
$d$-dimensional gauge invariance and with $O(\e)$ global invariance.
The Lagrangian is given by
\eqn\Ac{
L=L_{\rm SUSY}+L_{\rm SB}+L_{\e}.}
Here $L_{\rm SUSY}$ is the Lagrangian for  the $N=1$ supersymmetric
gauge theory, containing the gauge multiplet $\{A_{\m},\l\}$ ($\l$ being the
gaugino) and a chiral superfield $\Ph_i$ with component fields
$\{\ph_i,\psi_i\}$ transforming as a (in general reducible)
representation $R$ of the gauge group $\cal G$ (which we assume to
be simple). The superpotential is given by
\eqn\Aae{
W={1\over6}Y^{ijk}\Ph_i\Ph_j\Ph_k+{1\over2}\m^{ij}\Ph_i\Ph_j+L^i\Ph_i.}
The soft breaking is incorporated in $L_{\rm SB}$, given by
\eqn\Aaf{
L_{\rm SB}=(m^2)^j{}_i\ph^{i}\ph_j+
\left({1\over6}h^{ijk}\ph_i\ph_j\ph_k+{1\over2}b^{ij}\ph_i\ph_j
+ {1\over2}M\l\l+{\rm h.c.}\right)}
(Here and elsewhere, quantities with superscripts are complex conjugates of
those with subscripts; thus $\ph^i\equiv(\phi_i)^*$.)
Aside from the terms
included in $L_{SB}$ in
Eq.~\Aaf, one might in general have $\ph^2\ph^*$-type couplings,
$\psi\psi$ mass terms or $\l\psi$-mixing terms (as long as
they satisfy a constraint that quadratic divergences are not produced).
However, the
soft-breaking terms we
have included are those which would be
engendered by an underlying supergravity theory
(for a review, see \ref\sug{H.-P.~Nilles, \physrep  {\bf C}110 (1984) 1.})
and which are therefore
considered most frequently in the literature.

Finally,
$L_{\e}$ contains the evanescent couplings and is given by
\eqn\Ad{\eqalign{
L_{\e}&=g^2\ph^{*}R_A R_B\ph a_{\a,A}a_{\a,B}
+{1\over4}g^2f_{ABE}f_{CDE}a_{\a,A}a_{\b,B}a_{\a,C}a_{\b,D}
+{1\over2}\tm^2a_{\a,A}a_{\a,A}\cr
&+g(i\psib\s_{\a}R_A\psi + f_{ABC}\lambdab_B\s_{\a}\l_C ) a_{\a,A},\cr}}
where $a_{\a,A}$ are the $\e$-scalars and $f_{ABC}$ are the
structure constants of $\cal G$. The index $\a$ in $a_{\a,A}$
therefore runs from 0 to $\e$. The $\e$-scalars transform under the adjoint
representation of $\cal G$. There is another possible soft
term of the form $\xi^{ABi}a_{\a,A}a_{\a,B}\phi_i$ which
will be present if $\phi$ includes a representation
contained in the symmetric product of two adjoints, but we will ignore this
possibility for simplicity.

The non-renormalisation theorem tells us that the
superpotential $W$ undergoes no infinite renormalisation
so that we have, for instance
\eqn\Ada{
\b_Y^{ijk}=Y^{ijp}\g^k{}_p+(k\leftrightarrow
i)+(k\leftrightarrow j),}
where $\g$ is the anomalous dimension for $\Ph$.
The one-loop results for the gauge coupling $\b$-function $\b_g$ and
for $\g$ are given by
\eqn\Aab{
\lf\b_g^{(1)}=g^3Q \q\hbox{and}\q
\lf\g^{(1)i}{}_j=P^i{}_j,}
where
\eqn\Aac{
Q=T(R)-3C(G),\q\hbox{and}\q
P^i{}_j={1\over2}Y^{ikl}Y_{jkl}-2g^2C(R)^i{}_j.}
Here
\eqn\Aaca{
T(R)\d_{AB} = \Tr(R_A R_B),\q C(G)\d_{AB} = f_{ACD}f_{BCD} \q\hbox{and}\q
C(R)^i{}_j = (R_A R_A)^i{}_j.}
The two-loop
$\b$-functions for the dimensionless couplings were calculated in
Ref.~\ref\two{D. R. T. Jones, \npb87 (1975) 127\semi
A. J. Parkes and P. West, \plb138 (1984) 99;
             \npb256 (1985) 340\semi
             P. West, \plb137 (1984) 371\semi
          D. R. T. Jones and L. Mezincescu, \plb136 (1984) 242;
{\it ibid} 138 (1984) 293.},
and it was noticed that they can be written in the form
\eqn\Aad{\eqalign{ \llf\b_g^{(2)}&=2g^5C(G)Q-2g^3r^{-1}C(R)^i{}_jP^j{}_i\cr
\llf\g^{(2)i}{}_j&=-[Y_{jmn}Y^{mpi}+2g^2C(R)^p{}_j\d^i{}_n]P^n{}_p+
2g^4C(R)^i{}_jQ,\cr}}
where $Q$ and $P^i{}_j$ are given by Eq.~\Aac, and $r=\d_{AA}$.
It is therefore clear
that imposing the  conditions for vanishing of the one-loop
$\b$-functions for the dimensionless couplings, namely
\eqn\Aae{
Q=0\q\hbox{and} \q P^i{}_j=0 }
also guarantees that the two-loop $\b$-functions
vanish. One of our central results in this  paper will be that a similar
phenomenon happens in the case of the $\b$-functions for the soft
breaking couplings. Note that the conditions Eq.~\Aae\ imply that the
gauge group should have no $U(1)$ factors and that there should be no
gauge singlet fields.

The $\b$-functions for the physical soft-breaking couplings are most easily
computed
using the general results for DRED given in Ref.~\jo\ (being careful to
select the results corresponding to the first version of DRED
described above). The $\b$-functions for the $\e$-scalar mass, however,
must either be computed {\it ab initio} or else by coupling constant
redefinition starting from the results for DREG (see later for more details
on this process).
The one-loop $\b$-functions
for the soft-breaking couplings are given by
\eqn\Ac{\eqalign{
\lf\b_h^{(1)ijk}&=U^{ijk}+U^{kij}+U^{jki},\cr
\lf\b_b^{(1)ij}&=V^{ij}+V^{ji},\cr
\lf[\b_{m^2}^{(1)}]^i{}_j&=W^i{}_j+2g^2(R_A)^i{}_j\Tr[R_Am^2],\cr
\lf\b_M^{(1)}&=2g^2QM,\cr}}
where
\eqn\Aaa{\eqalign{
U^{ijk}&=h^{ijl}P^k{}_l+Y^{ijl}X^k{}_l,\cr
V^{ij}&=b^{il}P^j{}_l+{1\over2}Y^{ijl}Y_{lmn}b^{mn}
+\m^{il}X^j{}_l,\cr
W^j{}_i&={1\over2}Y_{ipq}Y^{pqn}(m^2)^j{}_n+{1\over2}Y^{jpq}Y_{pqn}(m^2)^n{}_i
+2Y_{ipq}Y^{jpr}(m^2)^q{}_r\cr &\quad
+h_{ipq}h^{jpq}-8g^2MM^*C(R)^j{}_i,\cr}}
with
\eqn\Aab{
X^i{}_j=h^{ikl}Y_{jkl}+4g^2MC(R)^i{}_j.}
The one-loop $\b$-function for the $\e$-scalar mass is given by
\eqn\Ah{
\lf\b_{\tm^2}^{(1)}=4g^2S+2g^2Q\tm^2,}
where
\eqn\Ai{
S\d_{AB}=(m^2)^i{}_j(R_AR_B)^j{}_i
-M\dM C(G)\d_{AB}.}
The two-loop
$\b$-functions for the physical soft-breaking scalar couplings
can be written in the following suggestive form:
\eqn\Ab{\eqalign{
\llf\b_h^{(2)ijk}&=-\Bigl[h^{ijl}Y_{lmn}Y^{mpk}+2Y^{ijl}Y_{lmn}
h^{mpk}-4g^2MY^{ijp}C(R)^k{}_n\Bigr]P^n{}_p\cr&\quad-2g^2U^{ijl}C(R)^k{}_l
+g^4(2h^{ijl}-8MY^{ijl})C(R)^k{}_lQ-Y^{ijl}Y_{lmn}Y^{pmk}X^n{}_p\cr
&\quad+(k\leftrightarrow i)+(k\leftrightarrow j),\cr
\llf\b_b^{(2)ij}&=\Bigl[-b^{il}Y_{lmn}Y^{mpj}
-2\m^{il}Y_{lmn}h^{mpj}-Y^{ijl}Y_{lmn}b^{mp}
\cr&\quad+4g^2MC(R)^i{}_k\m^{kp}\d^j{}_n\Bigr]P^n_p
-\bigl[\m^{il}Y_{lmn}Y^{mpj}+{1\over2}Y^{ijl}Y_{lmn}\m^{mp}\bigr]X^n{}_p
\cr&\quad-2g^2C(R)^i{}_kV^{kj}+g^2C(R)^i{}_kY^{kjl}Y_{lmn}b^{mn}\cr &\quad
+2g^4(b^{ik}-4M\m^{ik})C(R)^j{}_kQ+(i\leftrightarrow j),\cr
\llf[\b_{m^2}^{(2)}]^j{}_i
&=\biggl(-\Bigl[(m^2)_i{}^lY_{lmn}Y^{mpj}
+{1\over2}Y_{ilm}Y^{jpm}(m^2)^l{}_n+{1\over2}Y_{inm}Y^{jlm}(m^2)^p{}_l
\cr&\quad+Y_{iln}Y^{jrp}(m^2)^l{}_r+h_{iln}h^{jlp}\cr&\quad+
4g^2MM^*C(R)^j{}_n\d^p{}_i+2g^2(R_A)^j{}_i(R_Am^2)^p{}_n\Bigr]P^n{}_p\cr
&\quad+\bigl[2g^2M^*C(R)^p{}_i\d^j{}_n-h_{iln}Y^{jlp}\bigr]X^n{}_p
-{1\over2}\bigl[Y_{iln}Y^{jlp}+2g^2C(R)^p{}_i\d^j{}_n\bigr]W^n{}_p\cr
&\quad +12g^4M\dM C(R)^j{}_iQ+8g^4SC(R)^j{}_i\cr
&\quad +g^2\tm^2\Bigl[4g^2QC(R)^j{}_i-Y_{ikl}Y^{mkl}C(R)^j{}_m-
2Y_{ikl}C(R)^k{}_mY^{lmj}\Bigr]\biggr)+{\rm h.c.}\cr
\llf\b_M^{(2)}&=g^4\Bigl(8C(G)QM-4r^{-1}C(R)^i{}_jP^j{}_iM
+2r^{-1}X^i{}_jC(R)^j{}_i\Bigr)
,\cr}}
and the two-loop $\b$-function for the $\e$-scalar mass has the form
\eqn\Aj{\eqalign{
\llf \b_{\tm^2}^{(2)}\d_{AB}&=-2g^2W^j{}_i(R_AR_B)^i{}_j+8g^4C(G)S\d_{AB}
+8g^4C(G)QM\dM\d_{AB}\cr
&\quad-g^2\tm^2\Bigr(2Y_{ikl}Y^{jkl}(R_AR_B)^i{}_j
+{63\over2}g^2C(G)^2\d_{AB}\cr&\quad-8g^2C(G)T(R)\d_{AB}\Bigl),\cr}}
where $P^j{}_i$, $Q$, $W^i{}_j$, $X^i{}_j$
and $S$ are as defined in Eqs.~\Aac, \Aaa, \Aab\
and \Ai.
The results for the soft-breaking couplings agree with Refs.~\mv\ and \yam,
apart from $\b_{m^2}^{(2)}$. Our result for this does not even agree
with Refs.~\mv\ and \yam\ when we set $\tm^2=0$, since the term
$8g^4SC(R)^j{}_i$ is absent from the results given in these
references. We shall discuss this discrepancy in some detail later;
however we pause to note
a remarkable feature of the results which is at once apparent:
if we impose the
conditions Eq.~\Aae, together with the conditions
\eqn\Ak{
X^i{}_j=W^j{}_i=S=0,}
which guarantee that the one-loop $\b$-functions for the soft-breaking
couplings and the $\e$-scalar mass vanish\ref\jmy{D. R. T. Jones,
L. Mezincescu and Y.-P. Yao, \plb148 (1984) 317.}, and if furthermore
the $\e$-scalar mass is zero, then the two-loop $\b$-functions
for the soft-breaking couplings and the $\e$-scalar mass also vanish.
(We need to use the fact that $Q=0$ implies that $\cal G$ has
no $U(1)$ factors, which means that terms in $\b_{m^2}$ proportional
to $\Tr[R_Am^2]$ vanish, and also the fact that $P^i{}_j=0$ implies the
absence of gauge singlets which in turn guarantees $Y_{ijk}b^{jk}=0$.)
Hence the softly broken $N=1$ \sic\ theory
can be made finite up to two loops by imposing a very simple set of
conditions. Moreover, if the $\e$-scalar mass is chosen to be zero at some
scale, then in these particular circumstances the $\e$-scalars will not
develop a mass under running.

Now let us return to the question of the differences between
our results, and those obtained previously. The origin of the discrepancy
appears to be that the results of Refs.~\mv\ and \yam\ correspond to
the second version of DRED described earlier. As we mentioned earlier, this
version of DRED is not, in general, equivalent to DREG--an exception
being fully supersymmetric theories, for which in fact it gives the same
results as the version of DRED which we use. To support our contention that
the results of Refs.~\mv\ and \yam\ indeed correspond to this version of DRED,
and to investigate whether this version of DRED is viable for the
softly broken
\sic\ theory in the sense of giving results equivalent to those of
DREG, we need to explain the process of coupling constant redefinition in
more detail.
Let us start by considering the general case of a theory with
a set of couplings (both physical and evanescent) represented
generically by $\l^i$. We now imagine the $\e$-scalars to
have multiplicity $E$, rather than $\e$, and following Refs.~\hvd, \jjrb,
we consider the $\b$-functions
evaluated using DREG. If the $\b$-function $\b_{\rm DREG}^i$ for $\l^i$,
evaluated using DREG, has the form
\eqn\Af{
\b_{\rm DREG}^i(E)=\b_{\rm DREG}^i(E=0)+E\z^i+O(E^2),}
then up to two loops
the corresponding $\b$-functions $\b_{\rm DRED}^i$ evaluated using DRED
are obtained by making the coupling constant redefinitions
\eqn\Afa{
\l^i\rightarrow\l^i+\z^{(1)i}}
(where the superscript $(1)$ indicates the loop order) and have the form\hvd
\eqn\Ag{
\b_{\rm DRED}^i=\b^{(1)i}+\b_{\rm DREG}^{(2)i}(E=0)
+\b^{(1)j}{\del\over{\del\l^j}}
\z^{(1)i}-\z^{(1)j}{\del\over{\l^j}}\b^{(1)i}. }
(Note that at one loop, the DREG $\b$-function has no $E$-dependence and is
equal to the DRED $\b$-function.)
A fuller discussion of these results is given in Ref.~\jjrb.
Returning to the softly broken $N=1$ theory, we must be careful
when calculating the
$\b$-functions within DREG as described above, since the results obtained
using DREG are not manifestly supersymmetric. We should replace
$L_{\rm SUSY}$ by a general Lagrangian
for which we have not imposed the relations between the couplings
corresponding to supersymmetry. The two-loop $\b$-function for $m^2$ is then
given by
\eqn\Am{\eqalign{
[\b_{m^2}^{(2)j}{}_i]{}_{\rm DRED}&=[\b_{m^2}^{(2)j}{}_i]{}_{\rm DREG}
+\Bigl(\z_{\l}^{(1)}{\del\over{\del\l}}
+\z_{m^2}^{(1)l}{}_k{\del\over{\del (m^2)^l{}_k}}\Bigr)
\b_{m^2}^{(1)j}{}_i\cr
&\quad-\Bigl(\b^{(1)}_{\tm^2}{\del\over{\del\tm^2}}
+\b_{\l}^{(1)}{\del\over{\del \l}}\Bigr)\z_{m^2}^{(1)j}{}_i.\cr}}
Here $\l$ stands for all the couplings except for $m^2$ and $\tm^2$.
After performing the
derivatives with respect to the couplings as indicated in Eq.~\Am, one sets
all these couplings to their \sic\ values (so that
$[\b_{m^2}^{(2)j}{}_i]{}_{\rm DREG}$ can be calculated using the \sic\ values
for the couplings from the outset). It is
straightforward, using
\eqn\Ama{
\z_{m^2}^{(1)j}{}_i={g^2\over{8\pi^2}}\tm^2C(R)^j{}_i}
and Eq.~\Ah, together with the results for DREG for a general theory given in
Ref.~\jo, and the coupling constant redefinitions as given in
Ref.~\mva, to verify that Eq.~\Am\ does indeed reproduce the result for the
DRED $\b$-function for $m^2$ given in Eq.~\Ab. One can also easily check that
the result of Refs.~\mv, \yam\ for $[\b_{m^2}^{(2)j}{}_i]{}_{\rm DRED}$
corresponds to replacing $\z_{m^2}^{(1)^j}{}_i$
by $-\z_{m^2}^{(1)j}{}_i$ in Eq.~\Am, and then setting $\tm^2=0$. This
means that the results of Refs.~\mv, \yam\ are equivalent up to field
redefinition to those of DREG for zero $\e$-scalar mass. However, as we
argued earlier, if one is interested in the running of the couplings, the
$\e$-scalar will in general develop a mass and so one must be able to
compute $\b$-functions consistently when $\tm^2\ne0$. It is not
clear how to extend the calculations of Refs.~\mv\ or \yam\ away
from $\tm^2=0$
in a convenient fashion. One could simply define a new regularisation
scheme to be given by making the coupling constant redefinitions as in
Eq.~\Am, but with the opposite sign for $\z_{m^2}^{(1)j}{}_i$;
but this would lose all the advantages of DRED since one would be forced to
calculate with the general, non-supersymmetric theory using DREG before
making the coupling constant redefinitions. This
would be particularly awkward for actual $S$-matrix computations.
It would be preferable if one could specify a regularisation scheme at the
diagrammatic level
which would produce the same results as those obtained from the modified
version of Eq.~\Am. As it happens, the second version of DRED described
above does give the result for $\b_{m^2}^{(2)j}{}_i$ obtained in
Refs.~\mv, \yam. It is easy to see how one might be led to use this
second prescription in the present case, since this second version of DRED
only differs from ours in omitting a counterterm for the $\e$-scalar mass.
However, if one computes using this prescription for non-zero
$\tm^2$, one finds that it gives exactly the same result
for $\b_{m^2}^{(2)j}{}_i$ as Eq.~\Ab,
except for the absence of the term $8g^4SC(R)^j{}_i$--in other
words the $\tm^2$ terms are unaltered. This is $\it not$ the same as the
result of the modified version of Eq.~\Am\ for $\tm^2\ne0$. So it seems
there is no diagrammatic
prescription which reproduces the result of Refs.~\mv, \yam\
for $\tm^2=0$,
and which is equivalent to DREG for $\tm^2\ne0$. We believe that if one
wishes to perform the running coupling analysis one is ineluctably driven
to the prescription for DRED  which we have adopted.

We discovered above that a simple set of conditions sufficed to make the
full set of $\b$-functions for the softly broken
$N=1$ \sic\ theory vanish up to two loops. This result generalises what was
known already for the ordinary, unbroken $N=1$ theory. As we mentioned
earlier, we have only considered the soft-breaking terms which arise in
low-energy supergravity.
In fact, our results do not appear to
generalise to the case when other possible soft-breaking
terms, such as $\ph^2\ph^*$ interactions, are added; it then appears
that two-loop finiteness requires a completely
new set of conditions in addition to
those required for one-loop finiteness. It is very interesting that
finite softly broken \sic\ theories appear thus to be linked to the
existence of an underlying supergravity theory. In this context, we recall
that a solution to the conditions for one-loop finiteness of a softly broken
$N=1$ theory was found in Ref.~\jmy, namely
\eqn\An{
h^{ijk}=-MY_{ijk},\qquad (m^2)^i{}_j={1\over3}M\dM\d^i{}_j}
together with $Q=P^i{}_j=0$.
These relations are consistent with supergravity with
a minimal K\"ahler potential\sug\ (with the gravitino mass $m_{3\over2}$
given by $m_{3\over2}={1\over{\sqrt3}}M$). Remarkably, we find that the
solution Eq.~\An , together with $Q=0$,
also guarantees the vanishing of the one-loop
$\e$-scalar mass $\b$-function $\b_{\tm^2}^{(1)}$ given by Eqs.~\Ah, \Ai.
Thus Eq.~\An, together with $Q=P^i{}_j=\tm^2=0$, gives
a two-loop finite softly broken theory. Now in the \sic\ case, it
has been argued that from any two-loop finite theory one finite to
all orders can be constructed\ref\opz{R. Oehme, K. Sibold and W. Zimmermann,
                \plb 153 (1985) 142;
R. Oehme and W. Zimmermann  Com. Math. Phys 97 (1985) 569;
               W. Zimmermann, {\it ibid} 97 (1985) 211;
R. Oehme,  Prog. Theor. Phys. Suppl. 86 (1986) 215.}.
It is tempting to suppose that the same is true in the softly broken case.

Would such theories be of
phenomenological as well as theoretical interest? There has been
occasional interest\ref\jr{D. R. T. Jones and S. Raby, \plb 143 (1984) 137.}
\ref\utc{S. Hamidi and J. H. Schwarz, \plb 147 (1984) 301\semi
J. E. Bjorkman,  D. R. T. Jones and S. Raby, \npb 259 (1985) 503\semi
J. Leon et al  \plb 156 (1985) 66\semi
D. R. T. Jones and A. J. Parkes, \plb160 (1985) 267\semi
D. R. T. Jones, \npb 277 (1986) 153\semi
D. Kapetanakis, M. Mondragon and G. Zoupanos, \zpc 60 (1993) 181\semi
M. Mondragon and G. Zoupanos, CERN-TH.7098/93\semi
N. G. Deshpande, Xiao-Gang He and E. Keith, Oregon preprint OITS-540.}
in the phenomenology of finite $N=1$ theories, centering chiefly on a
particular $SU_5$ example first explored in Ref.~\jr. Now the low energy
effective field theory derived from such a theory is of course not
itself finite, because of the deletion of heavy fields. It would
therefore  appear that given that the unified model (the $SU_5$ one,
say) is in turn presumably a low  energy effective field theory then
there is no reason to  expect it to be finite. This is a compelling
argument, and forced in Ref.~\jr\ the  brave conjecture that the $SU_5$
model might indeed be the ultimate  theory, with gravitation
interactions being dynamically  generated. A less outr\'e possibility
is the following: evidently ``integrating out'' gravity from a theory is
very different from the process of integrating out the heavy fields
to produce a low energy theory; the graviton is after all massless.
Perhaps the nature of gravity is such that the resulting theory does
indeed remain finite. Clearly, however, substantive evidence for one or other
of these conjectures is required before we take seriously a postdiction for,
say, the top mass based on a finite model.
\vskip 10pt
\line{\bf Note Added \hfil}
We understand that the authors of Refs. [8] and [9] now agree that the DRED
result for $\b_{m^2}^{(2)j}{}_i$ is indeed given by our Eq. (16).
It is worth noting that (as has been independently observed by Martin and
Vaughn) the $\tm^2$ dependence in $\b_{m^2}^{(2)j}{}_i$ can be removed by a
simple redefinition of the form
$$ (\d m^2)^j{}_i=-{g^2\over{8\pi^2}}\tm^2C(R)^j{}_i.$$
The only other effect of this redefinition is to change the coefficient of
$g^4SC(R)^j{}_i$ in $\b_{m^2}^{(2)j}{}_i$ from 8 to 4.
It may well be that, as emphasised to us by Martin and Vaughn,
this modified scheme is the most convenient for
practical calculations.

We thank Steve Martin and Mike Vaughn for correspondence and for
drawing our attention to
a misprint in an earlier version of the paper.
\listrefs

\bye